\documentclass[a4paper,twocolumn]{esaproc} 
\newcommand{\Lone}{${\rm L}_1$}
\newcommand{\Ltwo}{${\rm L}_2$}
\usepackage{authordate1-4}
\usepackage{epsf}
\title{Space Debris Hazards from Explosions in the collinear Sun-Earth
Lagrange points}
\author[1]{M. Landgraf}
\author[1]{R. Jehn}
\affil[1]{ESA/ESOC, Robert-Bosch-Str. 5, 64293 Darmstadt, Germany}
\begin{document}
\maketitle
\begin{abstract}
The collinear Lagrange points of the Sun-Earth system provide an ideal
environment for highly sensitive space science missions. Consequently
many new missions are planed by ESA and NASA that require satellites
close to these points. For example, the SOHO spacecraft built by ESA
is already installed in the first collinear Lagrange point. Neither
uncontrolled spacecraft nor escape motors will stay close to the
Lagrange points for a long time. In case an operational satellite
explodes, the fragmentation process will take place close to the
Lagrange point. Apparently a number of spacecraft will accumulate
close to the Lagrange points over the next decades. We investigate the
space debris hazard posed by these spacecraft if they explode and fall
back to an Earth orbit. From our simulation we find that, as expected,
about half of the fragments drift towards the Earth while the other
half drifts away from it. Around $2\%$ of the simulated fragments even
impact the Earth within one year after the explosion.
\end{abstract}

\section{Introduction}
\begin{figure}[ht]
\center
\epsfbox{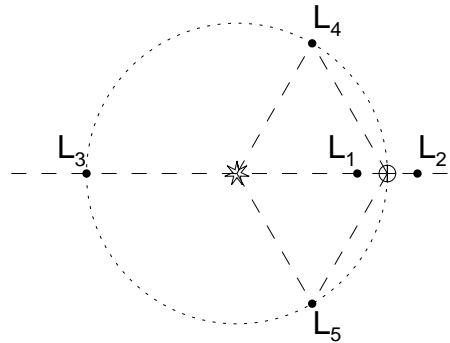}
\caption{\label{fig_points} Illustration of the position of the
Lagrangian points of equilibrium in the Sun-Earth system. The Sun is
at the centre of the plane representing the orbital plane of the
Earth, and the horizontal axis is fixed to the Sun-Earth line.}
\end{figure}

The hazards to manned as well as unmanned satellites by breakups of
satellites and rocket upper stages in low Earth orbit (LEO) and
geo-stationary orbit (GEO) have been extensively discussed in
the literature
\cite{jehn90,mcknight93,fucke93,jehn95,matney97,houchin97}. Since the
early 1970ies \cite{farquhar73} there are considerations to use
another class of orbits: quasi-stable trajectories around the Lagrange
points of the Sun-Earth system. Especially the collinear points \Lone\
and \Ltwo\ (see figure \ref{fig_points}) are interesting for solar
physics and space science applications \cite{farquhar98}. Since orbits
around the collinear libration points are inherently instable, the
dwell time of rocket bodies or uncontrolled satellites is short and
thus breakups of these objects are less likely than in LEO or
GEO. Breakups of malfunctioning operational satellites, however,
create a cloud of fragments on the stable manifold. The stable
manifold connects the locations of periodic motion in the
six-dimensional phase space of the satellite \cite{richardson80}.

In order to model the fragmentation process, we apply a fragmentation
model that combines the fragment mass distribution by Bess
\shortcite{bess75} with the $\Delta v$ distribution taken from the
explosion model by Reynolds \shortcite{reynolds90}. From these models,
the differential fragment mass distribution of a low intensity
explosion is given by (see also Jehn \shortcite{jehn90}):
\begin{eqnarray}
n d\!m & = & \left\{
\begin{array}{l}
1.71\times 10^{-4} M_t \exp(-0.02056\sqrt{m}) \\
\mbox{\hspace{5mm}for } m > 1936\:{\rm g} \\
8.69\times 10^{-4} M_t \exp(-0.05756\sqrt{m}) \\
\mbox{\hspace{5mm}for } m < 1936\:{\rm g}
\end{array} \right.
\end{eqnarray}
with $M_t$ being the total mass of the satellite. Assuming an average
fragment mass density of $\rho = 4.7\:{\rm g}\:{\rm cm}^{-3}$, the fragment
diameter $d$ is given by
\begin{eqnarray}
\frac{d}{2} & = & \sqrt[3]{\frac{3m}{4\pi\rho}}.
\end{eqnarray}
According to Reynolds, the $\Delta v$ of a fragment in $[{\rm m}\:{\rm
s}^{-1}]$ is given as a function of fragment diameter $d$ in $[{\rm m}]$ :
\begin{eqnarray}
\log\Delta v & = & -0.0676 \left( \log d \right)^2 - 0.804 \log d \\
 && - 1.514\nonumber
\end{eqnarray}

\begin{figure}[ht]
\center
\epsfbox{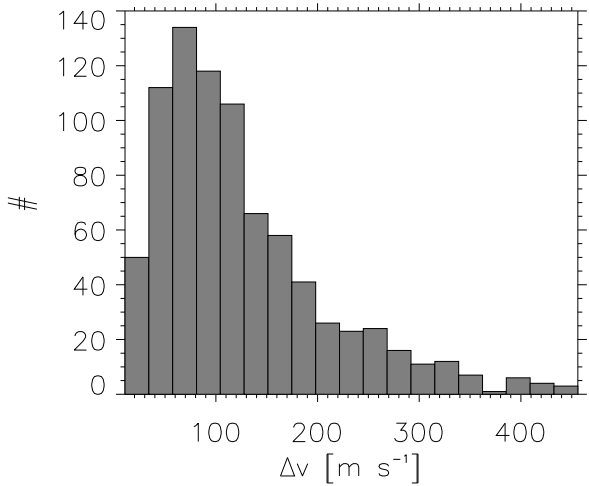}
\caption{\label{fig_deltavdis} Histogram of the distribution of
$\Delta v$ of the generated fragments (total number: $820$).}
\end{figure}

We simulate an explosion close to \Ltwo\ using the models described
above. A total number of $820$ fragments is generated, that drift off
the point of equilibrium depending on the direction and the magnitude
of the $\Delta v$ imposed on the fragment by the explosion. Figure
\ref{fig_deltavdis} shows the distribution of $\Delta v$ in the
simulated fragment cloud. We assume that the fragments are distributed
isotropically. After the simulated breakup we propagate the fragments
for one year, taking into account the gravity of the Earth, the Sun,
the Moon, as well as the oblateness of the Earth and solar radiation
pressure. In the following sections we describe the motion of the
simulated fragments, as well as the potential threat they pose to
operational satellites.

\section{Motion of the Fragments}
\begin{figure}[ht]
\vspace{2mm}
\center
\epsfbox{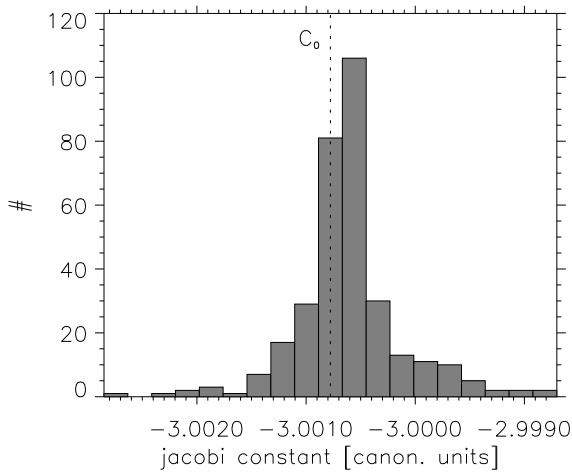}
\caption{\label{fig_jacdis} Histogram of the distribution of the
Jacobi constant of the generated fragments (total number: $820$). The
dotted line indicates the initial value of the constant before the breakup.}
\end{figure}

The dynamics of the fragments is dominated by the effective potential
in the Earth-fixed frame of the restricted three-body problem. Thus,
the fragment motion depends on the Jacobian constant that is
determined by the breakup process. Before the breakup, the satellite
is assumed to have the Jacobi constant of the \Ltwo\ point of
$C_0=-3.0008926$ in canonical units (distance unit $=1\:{\rm AU}$,
mass unit $=M_\odot + M_\oplus$, time unit such that period of Earth's
orbit equals $2\pi$). For this value of $C$, the satellite is at the
intersection of zero-velocity curves (ZVC), i.e., in equilibrium
motion. Since the satellite is in rest with respect to the Earth-fixed
frame, the breakup always increases the value of $C$. Because we
consider lunar gravity, the Earth's oblateness, and radiation
pressure, the Jacobi ``constant'' is not exactly constant along the
fragment's trajectories. The value of $C$ can be decreased by lunar
perturbations (e.g. by close fly-bys), or by the effect of the Earth's
oblateness during close Earth encounters. Figure \ref{fig_jacdis}
shows the distribution of $C$ for all $820$ fragments $30\:{\rm days}$
after the breakup.
\begin{figure}[ht]
\center
\epsfbox{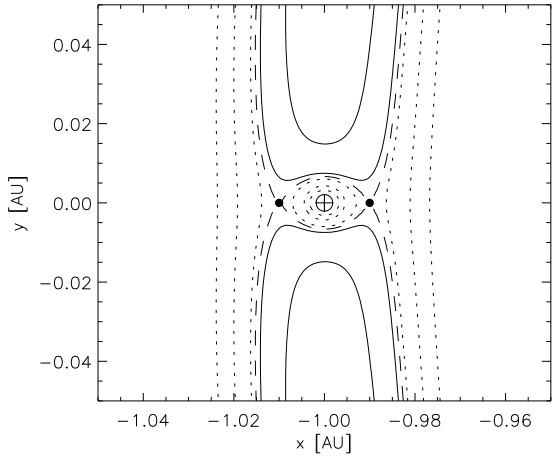}
\caption{\label{fig_zvc} Zero velocity curves (ZVCs) in the Earth-fixed
system for different values of the Jacobi constant. A plane
$0.1\times 0.1\:{\rm AU}$ around the Earth is shown, the Sun (not on
the diagram) is at the origin. The contour lines (solid for $C>C_0$,
dashed for $C=C_0$, dotted for $C<C_0$) represent ZVCs for $C=C_0,
C_0\pm 0.001, C_0 \pm 0.0005, and C_0 \pm 0.0001$.}
\end{figure}

\begin{figure*}[ht]
\center
 \epsfxsize=.8\hsize
\epsfbox{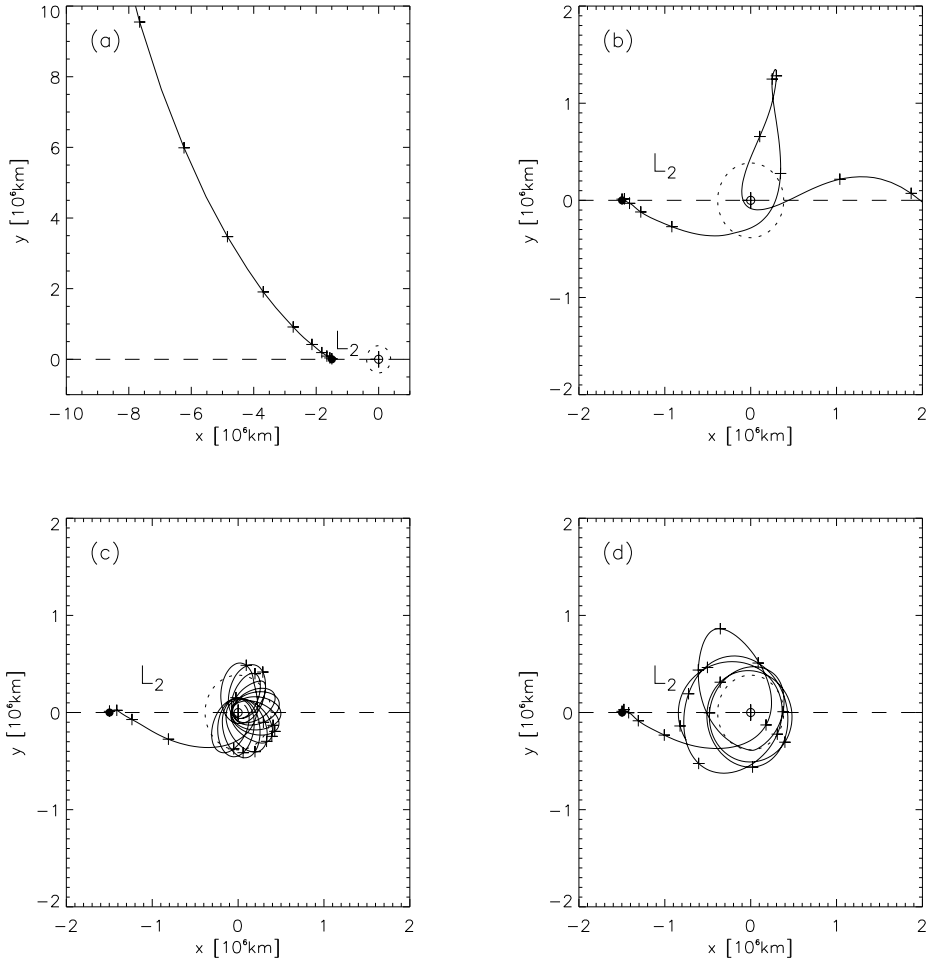}
\caption{\label{fig_traj} Trajectories (solid
lines) of an explosion fragments in the Earth's orbital plane. In
panel (a) a region $11\times 11\:{\rm AU}$, and in panels (b)-(d) a
region $4\times 4\:{\rm AU}$ are shown with the Earth at the
origin. The dotted circle indicates the orbit of the Moon and the
dashed line the Sun-Earth line. The Sun is located at $149\times
10^6\:{\rm km}$ along the positive $x$-axis. The cross-markers along
the trajectories indicate the fragment's position separated by $20\:{\rm
days}$. Panel (a) shows an Out-bound directed trajectory, panel (b) a
typical in-bound trajectory, and panels (c) and (d) in-bound
trajectories that are captured into bound orbits around the Earth.}
\end{figure*}

The breakup creates a rather narrow distribution of the Jacobian
constant around the equilibrium value $C_0$. Fragments with $C<C_0$
are confined either to a region close to the Earth or to outside a
ring around the Earth's orbit. This restriction can be seen in figure
\ref{fig_zvc} that shows the ZVCs for fragments with different values
of $C$. For $C>C_0$ the fragments may move freely along the Sun-Earth
line, but not along the Earth's orbit. In general two classes of
orbits are possible: out-bound and in-bound. Fragments on out-bound
orbits move away from the Earth so that they have larger heliocentric
semi-major axes and smaller mean motions. Consequently they drift into
the positive $y$-direction, which is opposite to the direction of
motion of the Earth around the Sun (see figure
\ref{fig_traj} (a)). These fragments, which constitute $44\%$ of
the simulated objects, continue on independent heliocentric orbits and
therefore pose no threat to Earth satellites. The other class of
orbits are in-bound to the Earth that potentially exhibit close lunar
or Earth encounters. If no close encounters occur, the orbits around
the Earth are very unstable, as shown in figure
\ref{fig_traj} (b). Mostly lunar fly-bys decrease the
orbital energy (and thus the value of $C$) of some fragments, so that
they stay in bound orbits about the Earth for more than one year (see figures
\ref{fig_traj} (c) and (d)).
\pagebreak

\section{Potential Hazards to Operational Satellites}
\begin{figure}[ht]
\center
\vspace{4mm}
\epsfbox{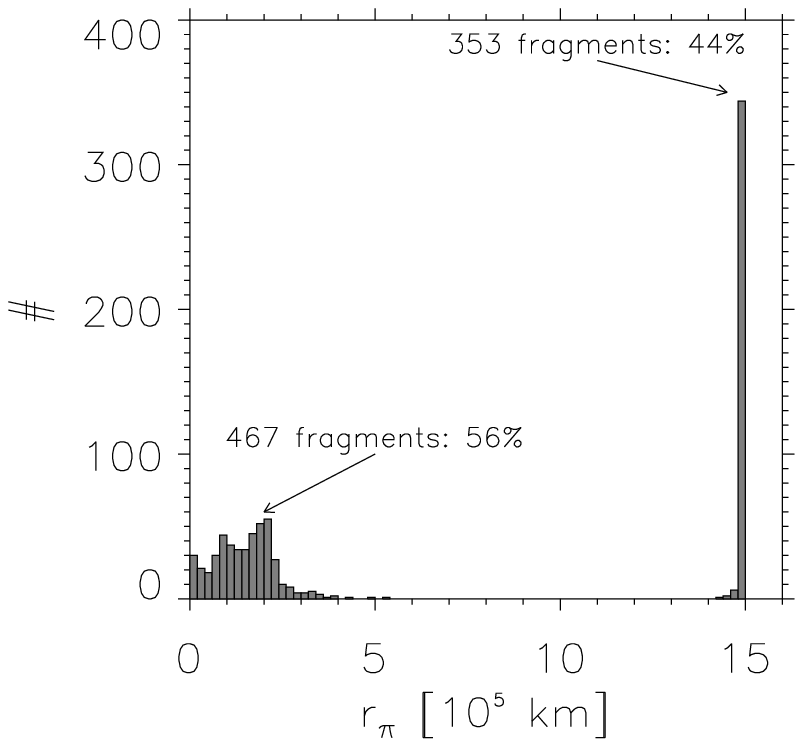}
\vspace{2mm}
\caption{\label{fig_perigeedis1} Histogram of the perigee distribution
of the fragments (total number: 820) one year after the breakup.}
\end{figure}
\begin{figure}[ht]
\center
\epsfbox{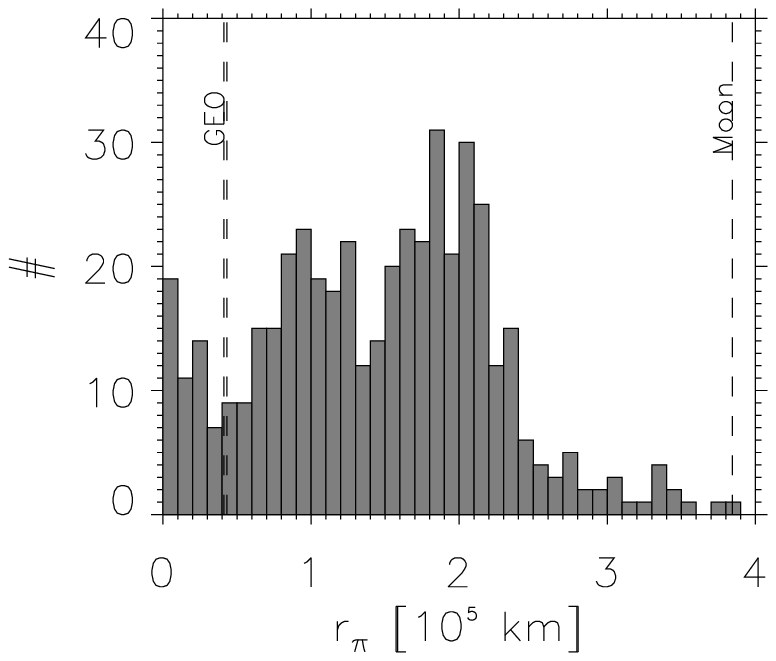}
\vspace{2mm}
\caption{\label{fig_perigeedis2} Zoom of the perigee range from $0$ to
$4\times 10^5\:{\rm km}$ in figure \ref{fig_perigeedis1}. The dashed
lines indicate the distance range of the GEO ring and the lunar
orbit.}
\end{figure}
In our simulation $56\%$ of the fragments moved towards the Earth
along the Sun-Earth line. Whether or not they pose a threat to
operational satellites depends on whether they reach LEO or GEO
distances, i.e. on their perigee distance $r_\pi$. Figure
\ref{fig_perigeedis1} shows the distribution of $r_\pi$ for all $820$
fragments, propagated over one year. The peak at $1.5\times 10^6\:{\rm
km}$ indicates the $353$ ($44\%$) fragments on out-bound orbits that never
approach the Earth closer than the distance of \Ltwo. The $467$
($56\%$) fragments on in-bound orbits almost always approach the Earth
closer than the lunar orbit. Figure \ref{fig_perigeedis2} shows a zoom
of the fragment perigee distribution inside the lunar orbit. The
maximum of the perigee distribution is at $2\times 10^5\:{\rm km}$,
far outside the orbit of most operational satellites. While less than
$60$ fragments approached the Earth closer than GEO, no fragment
actually intersected the GEO ring. In addition, a crossing of
fragments from \Ltwo\ of the GEO ring is only possible around the
equinoxes when the GEO ring intersects the Sun-Earth line.
\begin{figure}[ht]
\center
\epsfbox{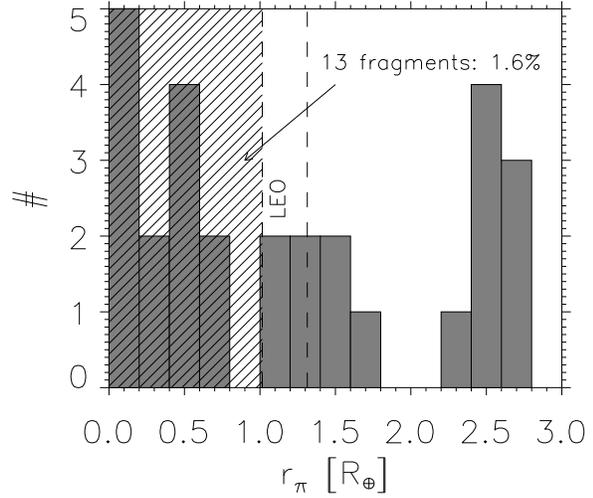}
\vspace{2mm}
\caption{\label{fig_perigeedis3} Zoom of the perigee range from $0$ to
$3 R_\oplus$ (Earth radii) in figure \ref{fig_perigeedis2}. The dashed
lines indicate the boundaries of the LEO region and the hashed area
indicates the extend of the Earth.}
\end{figure}

\begin{figure}[ht]
\center
\epsfbox{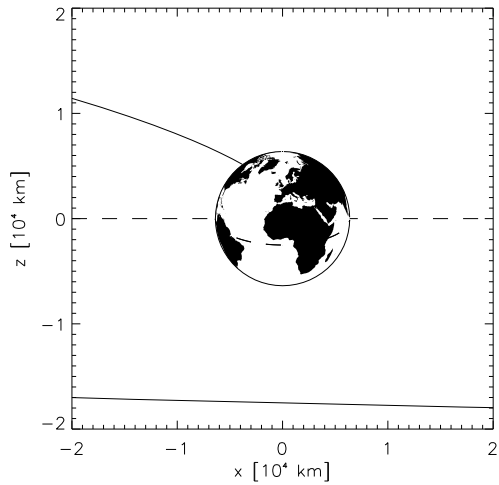}
\caption{\label{fig_earthimpact} Trajectory (solid line) of an
impacting fragment in the Earth's vicinity. The $x$-direction is the
Sun-Earth line and the $z$-direction is perpendicular to the Earth's
orbital plane. At the bottom a segment of the fragment's trajectory
from an earlier close encounter can be seen. The final segment of the
trajectory approaches the Earth from the upper left corner.}
\end{figure}

Due to their low initial angular momentum (or transversal velocity)
with respect to the Earth, fragments can actually impact the Earth's
surface. In our simulation $17$ fragments reached the LEO environment
($H<2000\:{\rm km}$), and $13$ ($1.6\%$) impacted the Earth close to
Earth escape velocity (see figure
\ref{fig_perigeedis3}). Despite the fact that the out-of plane motion
of the fragments stays small, they can impact the Earth at any
latitude, because the Earth's diameter small is compared to the distance
to \Ltwo (see figure \ref{fig_earthimpact}).

\section{Conclusion}
The use of the collinear Sun-Earth Lagrange points for solar physics
and space science application is expected to increase. Therefore, a
breakup of a controlled satellite due to a malfunction can not be
ruled out. Since the dwell time of uncontrolled satellites and rocket
bodies close to the libration points is short, they are unlikely to
contribute to the hazard to operational satellites due to breakups at
these points. If an explosion of a satellite occurs close to \Ltwo,
about $50\%$ of the fragments move in-bound towards the
Earth. Collisions of the fragments with GEO satellites are very
unlikely and are possible only if the breakup occurs close to the
vernal or autumnal equinox. A low percentage (about $2\%$) of the
fragments reaches the LEO environment and impacts the Earth's
atmosphere with velocities close to the Earth's escape speed of
$11\:{\rm km}\:{\rm s}^{-1}$. These Earth impacts can happen at any
geographic latitude and longitude.

\bibliography{debris,mas}
\bibliographystyle{authordate1}

\end{document}